# Dynamics of thermal and wetting footprint of a volatile droplet during Leidenfrost transition

Vikash Kumar

*Bernal Institute, School of Engineering, University of Limerick, Limerick, V94 T9PX, Ireland*

Vikash.kumar@ul.ie

## Abstract

The levitation of a volatile droplet on a highly superheated surface is known as the Leidenfrost effect. Wetting state during transition from full wetting of a surface by a droplet at room temperature to Leidenfrost bouncing, i.e., zero-wetting at high superheating, is not fully understood. Here, visualizations of droplet thermal and wetting footprint in the Leidenfrost transition state are presented using two optical techniques: mid-infrared thermography and wetting sensitive total internal reflection imaging under carefully selected experimental conditions, impact Weber number < 10 and droplet diameter < capillary length, using an indium-tin-oxide coated sapphire heater. The experimental regime was designed to create relatively stable droplet dynamics, where the effects of oscillatory and capillary instabilities were minimized. The thermography for ethanol droplet in Leidenfrost transition state (superheat range of 82K-97K) revealed thermal footprint with a central hot zone surrounded by a cooler periphery, indicative of a partial wetting state during Leidenfrost transition. High-speed total internal reflection imaging also confirmed the partial wetting footprint such that there are wetting areas around a central non-wetting zone. Result presented here using ethanol as a test fluid shed light on the geometry and dynamics of a volatile droplet footprint in Leidenfrost transition state.

**Keywords:** Leidenfrost transition, wetting dynamics, thermography, total internal reflection, droplet thermo-fluidics

## Introduction

Investigations on the dynamics of liquid droplet impingement on heated solid substrates and resultant wetting is of fundamental interest and the key to optimizing spray cooling thermal management strategies for electronics, as well for improving the performance of combustion engines[1–3]. The outcome of droplet impact on heated surfaces can be categorized into three regimes: *contact boiling*, *nucleate boiling* and *Leidenfrost boiling,* in order of increasing surface heating[4]. Understanding the wetting dynamics during transition from nucleate to Leidenfrost boiling could provide further insights into attaining the desirable extension of the Leidenfrost point to higher surface temperatures [5,6], thus enhancing the heat transfer behaviour between surface and droplet.

Wetting dynamics vary as surface temperature is increased: at low heater temperatures, full wetting contact boiling occurs, resulting in a circular disc-shaped footprint area forming between the droplet and heater surface; nucleate boiling is characterized by inhomogeneous circular wetting due to the presence of vapor bubbles; and the Leidenfrost regime is characterised by zero-wetting behaviour.

Several sophisticated experimental techniques have been used to understand the droplet dynamics on heated surface. Early efforts to study drop impact on heated surfaces was performed on opaque surfaces where visualization of the droplet footprint/wetting was not possible[7,8]. Moreover, some of the older studies relied on point temperature measurements, e.g. thermocouples, providing low resolution information on the thermal behaviour of this process. Later, photography was used with optically transparent surfaces to uncover wetting dynamics, however these techniques are highly inaccurate in discerning contact points, due to the transparent nature of the liquids[9]. Recently, non-intrusive temperature recording with thermal cameras and optical techniques, such as interferometry and total internal reflection (TIR), have been used for studying wetting dynamics during the Leidenfrost transition. Jung *et al*. analysed drop impact on a heated surface using simultaneous infrared (IR) thermography and high-speed photography, however no specific investigations on the superheats at which Leidenfrost transition occurs was performed[10]. Shirota *et al*. investigated a levitating droplet around the Leidenfrost point using the TIR technique and reported on the time- and length-scale of Leidenfrost dynamics[4]. Khavari et al. also used TIR with improved temporal resolution and found that the base of the droplet (i.e. footprint) is rather oscillatory in nature, for similar experimental conditions as used in Shirota et al[11]. Khavari et al. investigated droplet dynamics at We > 25 ($We = \frac{\rho v^2 d}{\sigma}$, where $\rho$ is density, v is impact velocity, d is droplet diameter and $\sigma$ is surface tension), however, Riboux et al. have pointed out that the threshold weber number for splashing (a fluid dynamics instability which results in disintegration of a droplet's edge into satellite droplets) is much lower for heated surface cases compared to room temperature impact and occurs at We ~ 50[12]. It is likely that the oscillations observed by Khavari et al is indicative of instability, i.e. splashing, hence investigations of drop impact at lower We to minimize these instabilities/oscillations and focus solely on Leidenfrost dynamics is needed.

Several other investigations were performed in the literature, where it was pointed out that instability exists for a droplet in the Leidenfrost regime when the size of the droplet is bigger than the capillary length (radius> 3×capillary length, where , where capillary length $\lambda = \sqrt{\gamma/\rho g}$ = 1.6 mm at room temperature for ethanol, $\gamma$ is surface tension, $\rho$ is density and $g$ is acceleration due to gravity) [13,14]. This effect is categorized as Rayleigh-Taylor instability and is marked by the appearance of vapor blisters or bursting bubbles. In a similar manner to the droplet base oscillations, such events caused by capillary instabilities can lead to droplet dynamics that are too complex to clearly investigate dynamic Leidenfrost transition.

The goal of the present paper is to experimentally investigate the thermal and wetting footprint of a droplet in the dynamic Leidenfrost transition regime. Of particular interest are the final moments of wetting transition from the nucleate to the Leidenfrost regime as the surface temperature approaches the Leidenfrost point, where the nature of droplet footprint needs to be clarified. To do this, droplet impact dynamics were investigated using mid-infrared and visible light TIR techniques, under carefully selected experimental conditions to minimize oscillatory and capillary instabilities; the ethanol droplets had limited radius ($R \sim 0.9$ mm and is 40% smaller compared to capillary length ($\lambda \sim 1.6$mm) for ethanol), and We was restricted to 7 (about an order of magnitude lesser than threshold We (~50) limit for splashing in the Leidenfrost regime reported by Riboux et al.[15]). Experiments were conducted on an Indium-tin-oxide (ITO) coated sapphire surface designed for infrared and TIR imaging, such that the spatially and temporally resolved thermal and wetting dynamics of droplets in the Leidenfrost transition state could be visualised.

## Materials and methods

A schematic of the experimental setup is presented in Figure 1. Three main elements of the setup are: (1) a droplet dispenser (World Precision Instruments), (2) a thermal camera (FLIR SC5600), and (3) a heater surface. The heater is made of two components (1) IR and white light transparent sapphire wafer, (2) IR opaque, white light transparent and electrically conductive ITO, which is emitter as well as a joule heating element. Two sizes and thicknesses of sapphire wafers were used in our experiments; a 100 mm diameter, 0.65 mm thick wafer for IR imaging and a 50 mm diameter, 2 mm thick wafer for TIR imaging. The thinner wafer resulted in background interference noise and hence was replaced with thicker sapphire for optical TIR imaging. The ITO layer was sputtered onto the sapphire wafer (Diamond Coatings UK). It had a 10 Ω/square sheet resistance and an AFM-measured RMS surface roughness of ~2 nm (AFM measured roughness is shown in Supplementary figure 1). The thickness of the ITO layer was ~700 nm, the rationale for using this thickness of ITO is that it blocks any IR radiation from the droplet dispenser side (see Figure 1a) while ITO itself acts as an emitter providing thermal information local to the impact surface.

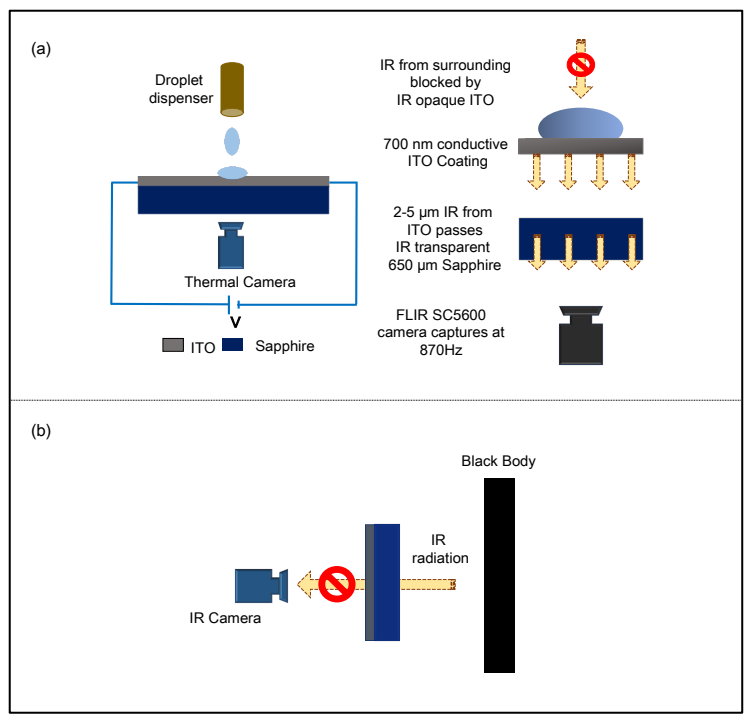

*Figure 1. Schematics of the experimental setup. (a) IR imaging experimental setup using a 26G stainless-steel needle to dispense droplets onto the heated surface and a backside IR camera to capture the thermal information. (b) IR opaqueness of ITO heater was confirmed by placing it in the path between a blackbody at a temperature (> room temperature) and an IR camera; camera did not read the blackbody temperature and rather indicated room temperature confirming opaqueness of ITO heater as it blocks IR coming from the blackbody.*

A confirmatory test was performed to confirm the IR opaqueness of ITO as shown in Figure 1b and details on optical properties of ITO are presented in Table 1 and extensive data over wide range of wavelength is presented in supplementary Table 1. Firstly, a blackbody was set at a temperature T (> room temperature) and its temperature was read using thermal camera as shown in Figure 1b. Then, the ITO heater was placed in the path between IR camera and the blackbody; temperature was read again using the thermal camera. In the second case, it was found that IR camera reads room temperature and not the temperature T of blackbody, hence confirming IR heater is an IR opaque material and blocks the IR radiation coming from the blackbody. A theoretical estimate of opaqueness of ITO could be made using Beer-lambert law, according to which the intensity of light decays exponentially with the distance and imaginary part of refractive index (k) could be used to calculate the penetration depth ($P_d$). The value of k at different wavelengths is provided in Table 1. $P_d$ gives an estimate of the distance at which intensity of the light falls to 1/e of original intensity, $P_d$ is defined below in equation (1).

$$P_d = \frac{\lambda}{4\pi k} \quad \text{...............................................................................................(1)}$$

An exemplary calculation is shown here; the value of k is 1.5 at wavelength of 4000nm and corresponding $P_d = \frac{4000nm}{4*3.14*1.5} = 212$ nm, which means light intensity would fall to 1/e at a depth of ~200nm. Thickness of ITO coating used in our setup is ~700nm, one would reasonably argue ITO behaves as an opaque surface in mid-IR region (3000-5000nm). The FLIR SC5600 thermal camera used in this paper responds best in mid-IR region as mentioned in the datasheet[16].

| Wavelength(nm) | n(real) | k(imaginary) |
|---|---|---|
| 520 | 2.09 | 0 |
| 580 | 2.06 | 0 |
| 662 | 2.01 | 0 |
| 770 | 1.93 | 0.01 |
| 1500 | 1.6 | 0.2 |
| 2000 | 1.4 | 0.5 |
| 3000 | 1.1 | 1 |
| 4000 | 0.9 | 1.5 |

*Table 1. Real and imaginary part of refractive index of ITO coating at various wavelengths*

During experiments, single spherical droplets of ethanol were released by slowly pushing the liquid through a 26G needle positioned at $h = 5$ mm above the desired test area on the heater surface for a range of surface temperatures. The instantaneous temperature of the heater surface in response to heat exchange with impinging droplet (i.e. thermal footprint) is captured with a thermal camera at a frequency of 870 Hz. The ITO-sapphire system was calibrated using a black tape (Type 88, 3M) of known emissivity of $\epsilon = 0.98$.

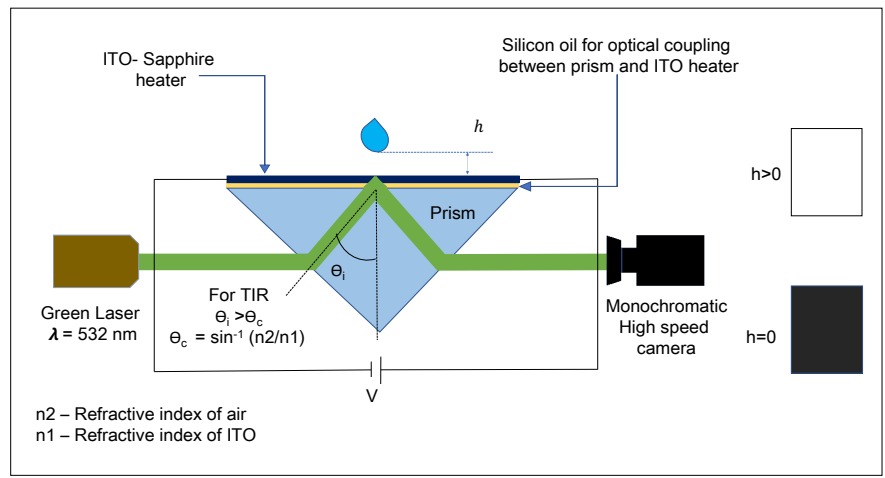

*Figure 2*. Schematic of experimental setup used for total internal reflection (TIR) imaging. Laser light guided by a prism undergoes total internal reflection at the ITO-air interface. Changes in optical transmission at the ITO surface are captured by a monochromatic high-speed camera. Depiction of high-speed camera pixel greyscale value changing from white to black as the droplet interface contacts the heated surface.

To track droplet wetting on the heater surface, we implemented TIR high-speed imaging[4]. As shown schematically in Figure 2, the TIR setup has four main components: (1) a $\lambda = 532$ nm green laser source, (2) a right angle prism of NBK-7 glass, (3) the ITO-sapphire heater surface placed on top of the prism with silicon oil (optical refractive index ~ 1.4) sandwiched in between the two to facilitate optical transmission, (4) a high speed camera (Phantom v311, Vision Research Inc.) with a frame rate up to 0.4 MHz, ~460X faster than the IR frame rate. The laser light was directed onto the prism at an angle $\theta_i$ greater than $\sin^{-1}(n_{air}/n_{ITO})$ such that the light would undergo TIR at the ITO-air interface, but less than the critical angle $\theta_c$ for the ITO-liquid interface, $\theta_c < \sin^{-1}(n_l/n_{ITO})$. This condition ensures that when the droplet lands on the heater, the white pixels (light reflected back from ITO-air interface) change to black (light transmitted into droplet from ITO), thus providing a large contrast image of the droplet wetting footprint.

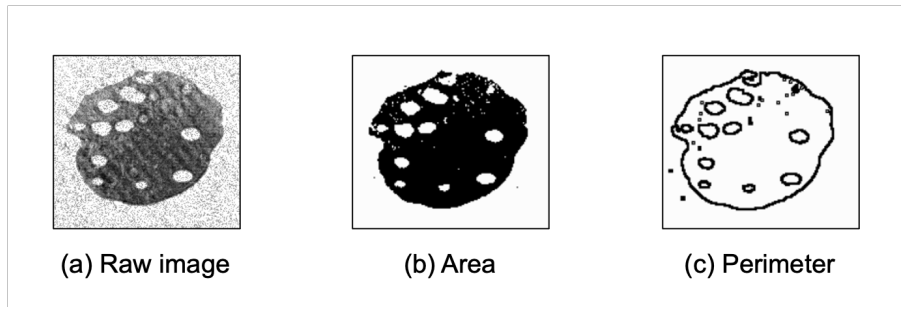

(a) Raw image  (b) Area  (c) Perimeter

*Figure 3. Post processing of TIR raw images (a). Raw image was converted into binary format; counting the number of black pixels in binary format yields area (b). Subsequently, edge tracking was done and black pixels was counted to deduce the perimeter of footprint (c).*

IR images were captured and analysed in FLIR Altair software. TIR images were post processed using ImageJ and subsequently analysed in MATLAB by first performing a required background subtraction on the acquired images to isolate the experimental features and then converted them to binary as shown in Figure 3 as an example. The raw image was captured at surface temperature of 135ºC as shown in 3a. For calculating the area, we counted the number of black pixels in the binary image seen in 3b. Image was further processed with in-built edge tracking library of ImageJ and another binary image was obtained as seen in 3c. Counting the black pixels in this edge-tracked image yields perimeter of the footprint.

# Results and discussions

Figure 4a and 4b show the spatial details of the thermal and wetting footprint of ethanol droplet captured using the IR and TIR setup shown in Figure 1 and 2 at various surface temperatures ($T_s$). A schematic of droplet dynamics expected from side view when looking at a plane perpendicular to heater surface and passing through centre of droplet footprint is shown in Figure 4c based on the insights gained from TIR and IR analysis. At $T_s$ = 150 °C [superheat $\Delta T = (T_s - T_{bp})$ = 72 K], above the nucleation onset superheat of $\Delta T$ = 37 K, a non-uniform, temporally fluctuating thermal footprint was observed, ~10 K cooler than the surrounding surface indicating the presence of small nucleating vapor bubbles underneath the droplet. Expectedly, TIR images confirms several small white patches where the vapor bubbles are present on the surface. At $\Delta T$ = 92 K, it was found that the thermal footprint exhibits a "dimple" of higher temperature beneath the centre of the droplet. The appearance of this dimple is attributed to the presence of a central trapped vapor (produced as a result of phase change) underneath the droplet, while the periphery is still in contact with the heater. This is supported by the wetting footprint shown in Figure 4b, where a central non-wetting zone is visible. The peripheral contact around the dimple, also visible in the TIR images in Figure 4, facilitates heat transfer and registers in the IR image as 12 K cooler than the surrounding surface and 3-4 K cooler compared to the hotter central region where heat transfer is comparatively inhibited due to a central trapped vapor. At a higher superheat of $\Delta T$ = 112 K, a uniform thermal footprint 4-5 K cooler than the surrounding surface was observed, indicating that the droplet was levitating on a vapour layer i.e., Leidenfrost state. No contact with the heater surface was seen at $\Delta T$ = 112 K with the TIR, as Figure 4(b) does not show any black pixels on the camera sensor. Arising from the experimental observations, three different boiling regimes can be defined: (1) *nucleate boiling* (2) *transition boiling* and (3) *Leidenfrost* boiling, where the transition from a non-uniform to uniform thermal footprint, with increasing superheat temperature, was demarcated by the formation of thermal dimples in transition boiling state in a temperature window of ~82-97K.

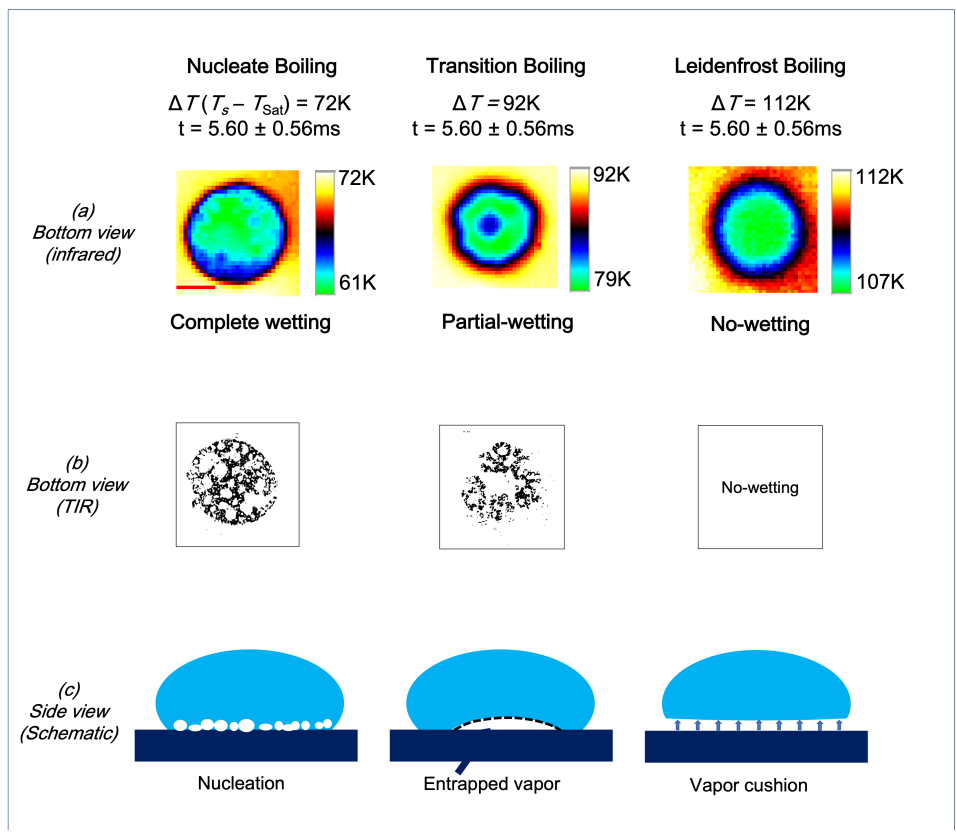

*Figure 4* Spatial details of (a)thermal and (b)wetting footprint for experiments with a 2 mm diameter ethanol droplet impacting the heater surface at We = 7 for various surface temperatures. (c)The anticipated characteristic side view of the impact event at the various superheats is shown schematically. Scale bar equals 1mm.

Figure 5 (a) and (c) show the experimental images of short-term temporal evolution (up to 10ms after first contact) and long-term evolution (up to 156ms after first contact between droplet and heater) of thermal dimples observed at $\Delta T$ = 92 K shown in Figure 4. Footprint of the droplet is stable temporally at We=7 compared to Khavari et al. observation of time dependent oscillatory footprint at We 30-300 which may be linked to splashing/oscillatory instability. It has been pointed out by Ribaux et al. theoretically that threshold We for splashing in Leidenfrost state could as low as ~50[12]. Figure 5 (a) shows the short-term evolution (within 6.72 ± 056ms of impact) of thermal footprint in the transition regime at superheats of $\Delta T$ = 92 K, a thermal dimple is observed at the center of the droplet footprint. The droplet spreads gradually on the heater surface with time. It is to be noted that droplet bounces off the surface at $t$ = 19.60 ± 0.56ms ($t/\tau$ = 3.3 ± 0.07) as seen in the long term temporal evolution tracked upto 156 ± 0.56ms after impact, where $\tau$ is defined as contact time of a bouncing drop based on inertio-capillary dynamics of the droplet ($\tau = \sqrt{\frac{\rho R^3}{\sigma}}$, where $\rho$ is density of liquid, R is radius and $\sigma$ is surface tension)[17]. It is seen that dimple remains centrally positioned before drifting out to the periphery as the droplet rebounds from the surface at $t$ = 19.60 ± 0.56ms. Later the droplet exhibits Leidenfrost boiling behavior as shown in thermal footprint at $t$ = 53.26 ± 0.56ms and eventually rolls out to the side as shown moving out of the frame at $t$ = 156.04 ± 0.56ms. The heater surface remains hot enough following the initial droplet rebound to induce sufficient vapor pressure to keep the droplet levitated. Also, the droplet might already be at a temperature>room temperature so that during its subsequent impact it transitions to a levitation stage. It is to be noted we also saw touch-down/contact in some of the experiments done in 82K<$\Delta T$<97K during 2nd or subsequent impact while some experiments yielded levitation reported above. Further discussions on footprint during 2nd impact is beyond on the scope of this report, however it is confirmed that footprint remains stable in any event.

The temperature profile, expressed in terms of superheat, along the diametric line was recorded (as shown with a red bi-directional arrow in Figure 5a) and plotted against position normalized by diameter of the droplet in Figure 5b. The periphery was observed to be cooling with time due to heat exchange with the surface; it was ~15 K cooler, while the central region of the thermal dimple footprint cools by 2-3 K, 6.16ms after impact.

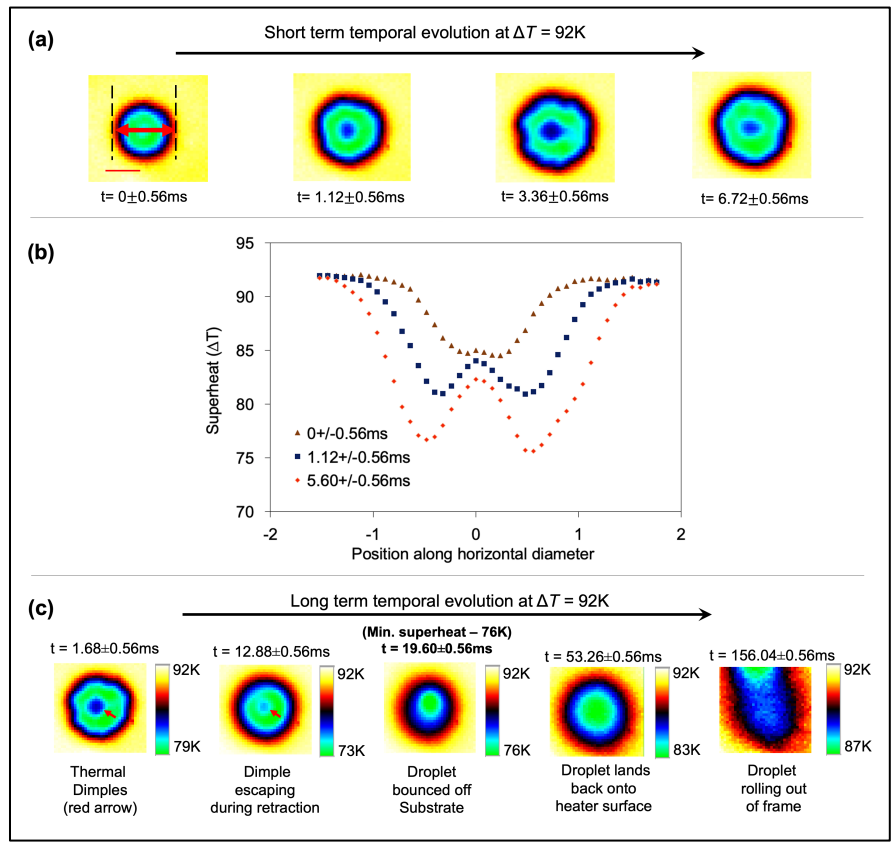

*Figure 5. Temporal evolution of thermal dimples (a) Thermal images showing short term temporal evolution of footprint at ΔT = 92 K (b) Plot of superheat against normalized position along diameter at three-time instance. Scale bar equals 1mm. (c) Thermal images showing long term temporal evolution (up to 156.04 ± 0.56ms after impact) of footprint at ΔT = 92 K.*

Figure 6 shows a side-view schematic with a central trapped vapor of height h. Estimation of h can made from noting that T1=T2 before impact, as the experimental setup was allowed to reach steady state and heat transfer on both sides was only possible via ambient air convection and radiation to the ambient. An estimate of the thermal diffusion length ($l_{thermal} = \sqrt{\alpha t} \sim 200 \mu m$, where $\alpha$ is thermal diffusivity), $l_{thermal} > y$ (thickness of sapphire wafer, 0.65mm) suggests that T1 before impact and at time t = 6.72 ± 056ms after impact should be approximately the same. Considering the sapphire wafer had a diameter of 100mm, which was much greater than its 0.65mm thickness, 1D steady state heat transfer is assumed and used to estimate the amount of heat conducted through the sapphire wafer.

$$Q'' = K_{sapphire} \frac{T_1 - T_2}{y} \dots\dots\dots\dots\dots\dots\dots\dots\dots\dots\dots\dots\dots\dots\dots\dots\dots\dots\dots\dots\dots\dots\dots\dots\dots\dots\dots\dots\dots\dots\dots\dots\dots\dots\dots\dots\dots(2)$$

Further, Q" is also the heat transferred to the liquid droplet through the vapor gap of height h. Since the lateral dimension of droplet footprint, as shown in Figure 3(left), is about 1mm >> h (O(10$\mu m$)), one can use 1D steady state heat transfer to calculate h using Q" obtained from (2).

$$Q" = K_{air} \frac{T_2 - T_{sat}}{h} \quad \dots\dots\dots\dots\dots\dots\dots\dots\dots\dots\dots\dots\dots\dots\dots\dots\dots\dots\dots\dots\dots\dots\dots\dots\dots\dots\dots\dots\dots\dots\dots(3)$$

Using (2) and (3), the estimate of the dimple height h ~10$\mu m$. To support this calculation, interferometric measurements of the vapor layer thickness in the Leidenfrost regime [Superheat = $\Delta T$ = 112 K] at We=7 underneath the ethanol droplet were also performed (experimental setup and image captured is shown in Supplementary Figure 2) and it was found to be O(1)µm.

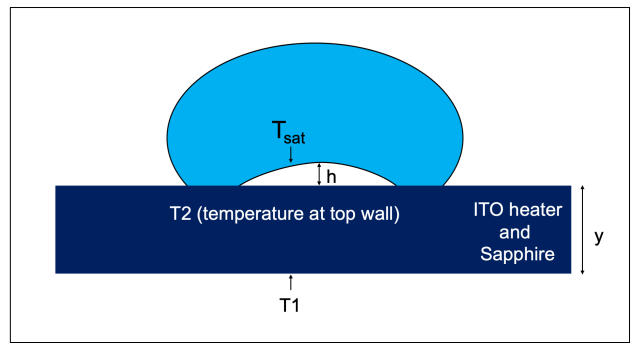

*Figure 6. Schematic of side-view of a droplet in transition regime showing the thermal boundary values for dimple height h calculations.*

The estimate of vapor height of the dimple in the transition regime obtained here could then be compared with the vapor height of a droplet in the Leidenfrost regime reported in the literature[18–20]. For example, Burton et al. used high resolution side view imaging to estimate the equilibrium vapor height underneath a water droplet for the case of static Leidenfrost at a surface temperature of 370°C[19]. It was found to be in the range of 10-100$\mu m$, while Biance et al. also report the vapor height to be about ~20$\mu m$ for the static case[13,19]. Tran et al. employed interferometry to estimate the vapor thickness underneath a water droplet at We=3.5 and 350°C during dynamic Leidenfrost impact and its central thickness was found to be about 2-3$\mu m$[21]. Thus, observations reported here and estimates of vapor height (h) are of the same order of magnitude as reported in the literature.

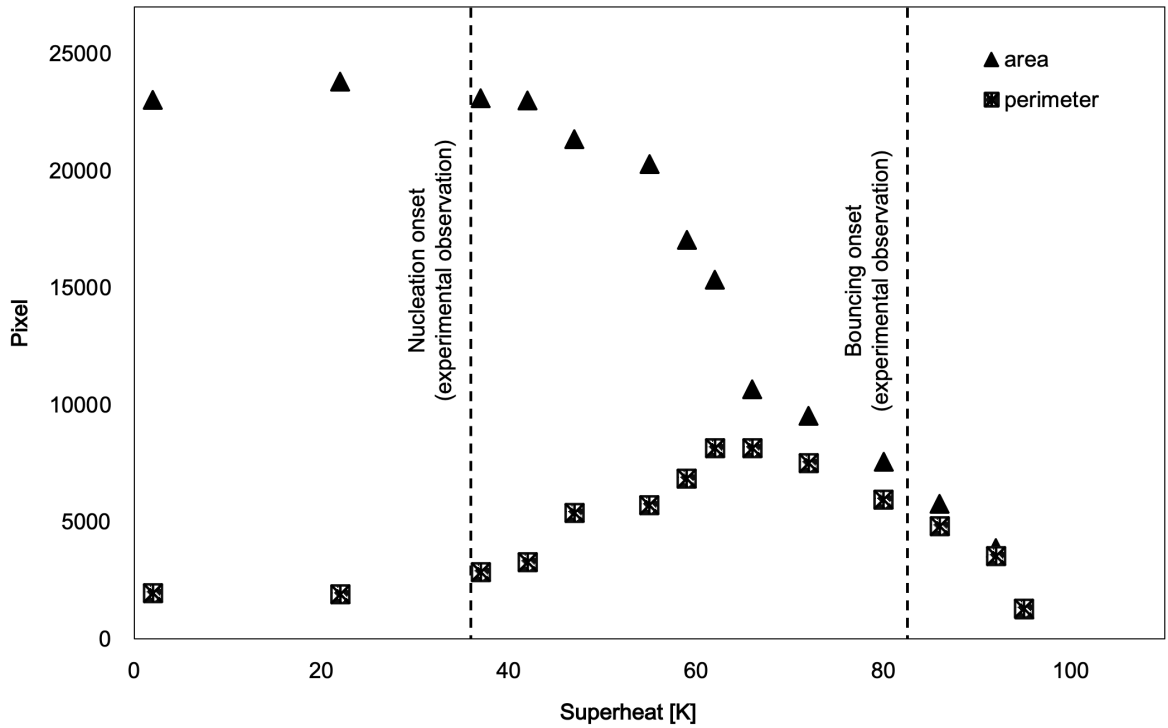

*Figure 7. Plot of the area and perimeter of wetting footprint in camera pixel units*

Next, we analysed the wetting footprint of the ethanol droplet captured with TIR technique. The nucleation onset ($\Delta T = 37$ K) and bouncing onset ($\Delta T = 82$ K) has been highlighted with vertical broken line as a visual reference. Because the transition regime represents reduction in the wetting footprint, the trend in both wetting area and perimeter (in pixel units) were examined at various superheats through the transition regime and plotted in Figure 7. These values were measured at the instant of maximum lateral extension of the droplet. We found that the wetting contact area monotonically decreases with increasing surface superheating, reducing by ~83% from heater surface temperatures close to the normal boiling point to a superheat of $\Delta T = 92$ K. This trend of reduction in wetting area can be understood by considering that the heater becomes progressively covered by vapor bubbles as superheat increases. We also plotted the perimeter, and it was found to be increasing as the droplet enters into nucleate boiling regime from contact boiling regime. Increase in number of nucleating bubbles produced on the surface as a result of phase change leads to increase in total triple contact line length at liquid-vapor interface and consequently leading to ~4-fold increase in perimeter as $\Delta T$ is increased from 0K to 64K. Above 64K superheating, the perimeter was seen to be decreasing which can be attributed to increase in vapor content and decrease in wetting footprint and hence leading to decrease in total amount of area available for nucleation.

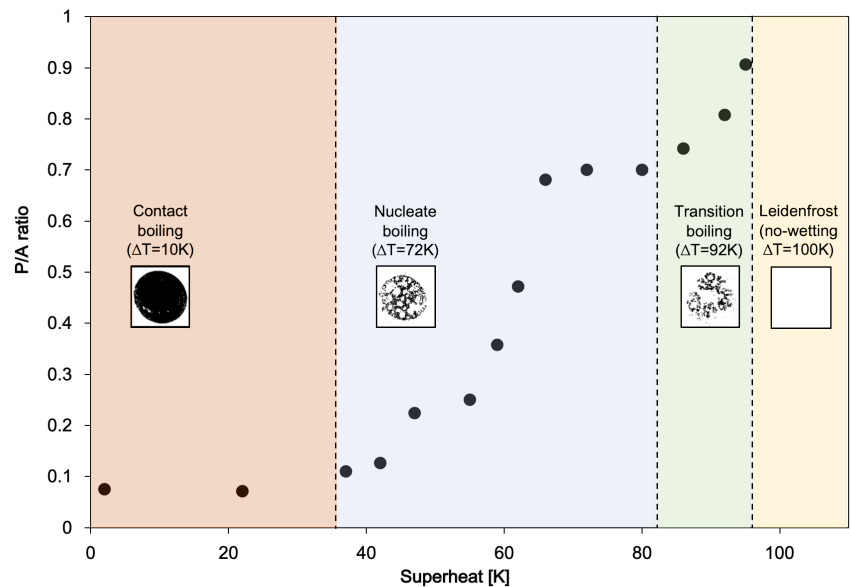

*Figure 8.* Plot of the P/A ratio at various superheats. Inset figures show the wetting footprint as a representative of various regime of boiling.

Further we plotted perimeter/area (P/A) ratio in Figure 8 at various superheat with an intent to demarcate various regime of droplet boiling. P/A ratio was found to be varying monotonically; it is possible that temperature dependent decrease in surface tension of ethanol-air interface allowed creation of extra ethanol-air interface per unit area which otherwise is not observed at room temperature. For example, the surface tension at ethanol-air interface at room temperature is ~22.4 mN/m and it decreases by ~30% to 15 mM/m at 373K[22]. The contact boiling regimes belongs to a regime of low value of P/A (<=0.1), while nucleate boiling regime corresponds to moderate value of P/A (0.1-0.7) where the onset of nucleate boiling regime was identified based on visual observation of nucleating bubbles seen in captured TIR footprint. The Leidenfrost or bouncing regime maps to high values of P/A > 0.7, and the boundary of Leidenfrost regime was identified by visual observation of bouncing of droplet seen on the surface. This classification scheme is monotonic and quantitative compared to qualitative schemes of using secondary atomization as an indicator of Leidenfrost onset or non-monotonic evaporation time[6,18]. The secondary atomization might be induced by other mechanisms such as splashing instabilities, while the evaporation time (lifetime of droplet on the heater) is susceptible to errors in time estimation due to practical reasons, especially in the Leidenfrost regime where the droplet is too mobile to be held in place to estimate it's lifetime and there is inevitable mass-loss due to thermal atomisation/bubble bursting in nucleate boiling regime[23].

# Conclusions

In conclusion, it was found that an ethanol droplet exhibits a spatial partial wetting state in transition state from nucleate boiling to Leidenfrost boiling when the experiments were performed with droplet radius ~ 0.9mm and We was restricted to 7. The wetting footprint was tracked over a long period of time and found to be stable. Height of vapor in central non-wetting zone, deduced from thermography, found to be $O(1)\mu m$ is in agreement with data reported in literature on vapor height in Leidenfrost state. TIR imaging provided the geometric details of the wetting footprint and it is seen that P/A ratio of the footprint increases monotonically with the surface temperature. It is further discussed that various stages of droplet boiling could be quantitatively categorized using P/A as a parameter. In light of the results presented here, effect of varying surface energy on the footprint geometry is currently being investigated and findings will be discussed elsewhere.

**Acknowledgement**

This research project has received funding from the European Union's Horizon 2020 research and innovation programme under the Marie Skłodowska-Curie grant agreement No. 643095.

# Supplementary

## AFM image of ITO coating

The AFM measured roughness of a 10 ohm/sq coating is ~ 2nm. Both the ITO coating and roughness measurement was performed by Diamond coatings UK.

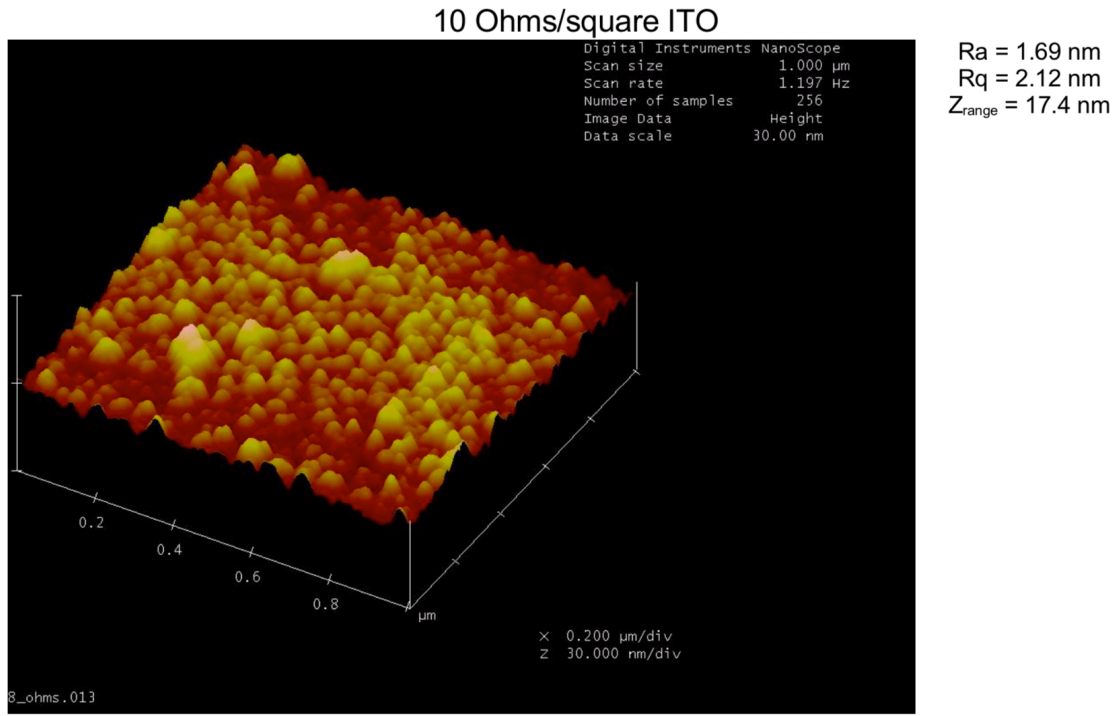

*Supplementary Figure 1. AFM measurement of ITO surface*

# Optical data of ITO coating

| Wavelength(nm) | n(real) | k(imaginary) |
|---:|---:|---:|
| 300 | 2.1 | 0.6 |
| 325 | 2.1 | 0.3 |
| 350 | 2.1 | 0.15 |
| 375 | 2.1 | 0.06 |
| 400 | 2.1 | 0.02 |
| 437 | 2.1 | 0.01 |
| 470 | 2.1 | 0.01 |
| 520 | 2.09 | 0 |
| 580 | 2.06 | 0 |
| 662 | 2.01 | 0 |
| 770 | 1.93 | 0.01 |
| 910 | 1.83 | 0.01 |
| 1500 | 1.6 | 0.2 |
| 2000 | 1.4 | 0.5 |
| 3000 | 1.1 | 1 |
| 4000 | 0.9 | 1.5 |
| 5000 | 0.8 | 2 |

*Supplementary Table 1. Data 'n' (real part of refractive index) and 'k' (imaginary part of refractive index) supplied by Diamond coatings UK.*

# Interferometry setup

A setup was built to measure the vapor gap using interferometry as shown in Supplementary Figure 2a. The interference between light reflected from top surface of heater and bottom surface of droplet leads to formation of fringes as shown schematically in Supplementary Figure 2b. A crude calibration method was developed to estimate the order of magnitude of thickness of vapor gap; details of which could be found Figure 2 of the reference (van der Veen et al)[1]. Briefly, a lens of known curvature was placed onto the setup and interferometric fringes were captured. The intensity variations as a function of spacing between lens and surface was plotted to make a calibration chart, it was used as a reference to estimate the vapor height between droplet and surface. For the images shown in Supplementary figure 2c, we found that vapor height is O(1) µm. An improved calibration method is under development and would be used to extract the complete vapor height profile in future.

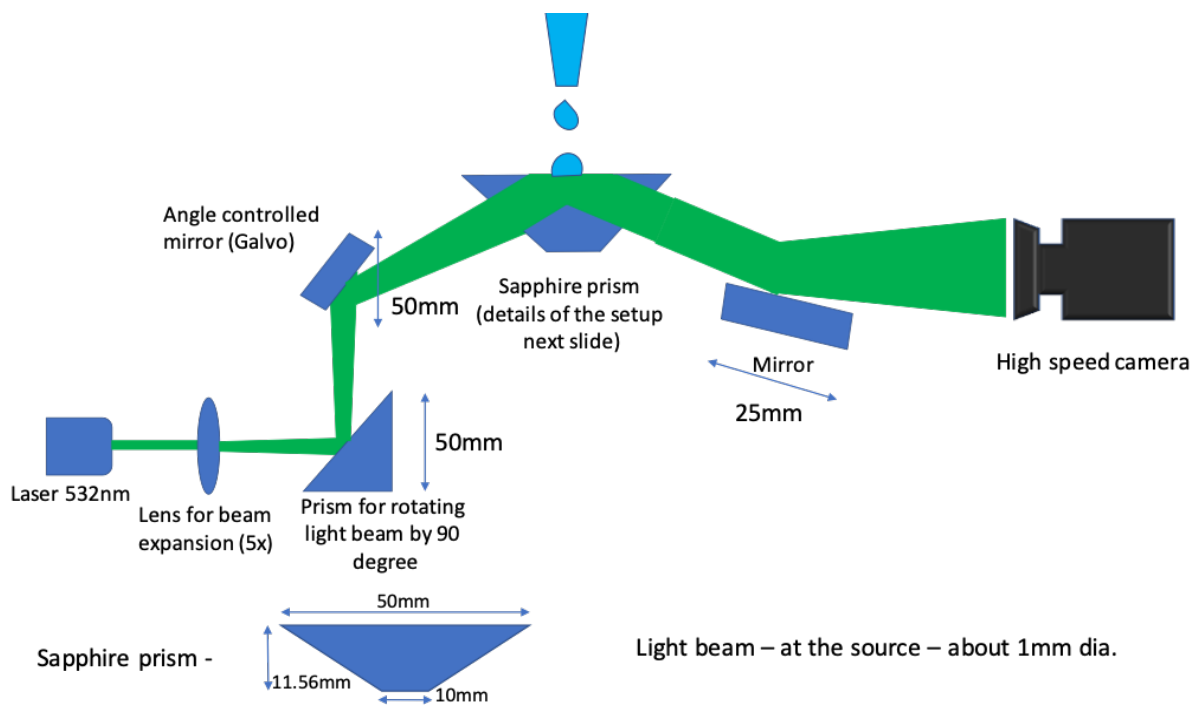

Supplementary Figure 2a. Experimental setup built for interferometric measurement of vapor gap between droplet and surface.

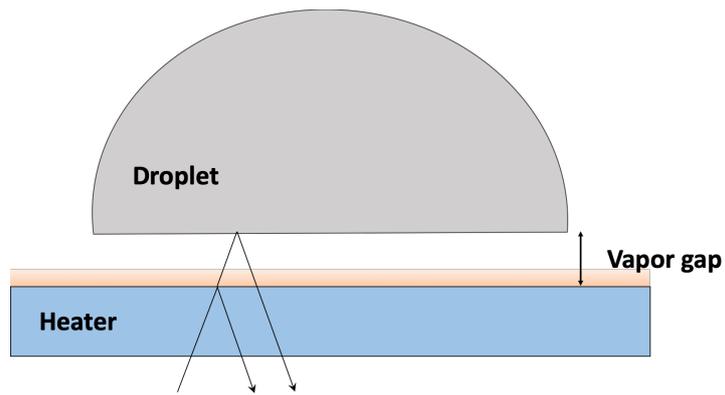

*Supplementary Figure 2b. Light reflected from top surface of heater and bottom surface of droplet leads to formation of fringes as shown in Supplementary figure 2c.*

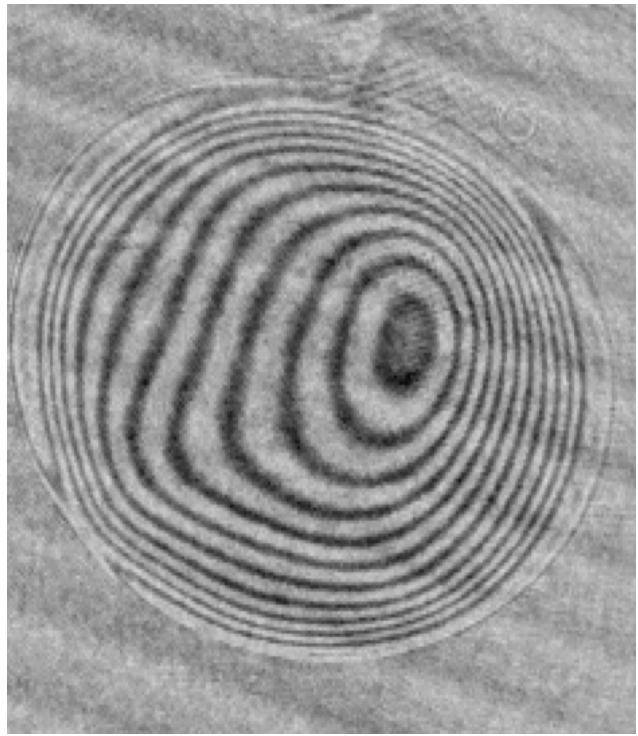

*Supplementary Figure 2c. Interferometric pattern captured using above-described setup at We~7 and ΔT=112K.*

*References:*

1. Van Der Veen RCA, Tran T, Lohse D, Sun C. Direct measurements of air layer profiles under impacting droplets using high-speed color interferometry. *Phys Rev E - Stat Nonlinear, Soft Matter Phys*. 2012;85(2):1-6. doi:10.1103/PhysRevE.85.026315.